\documentclass{llncs}
\usepackage{graphicx}
\usepackage{amsmath}
\newcommand{\be}{\begin{itemize}}
\newcommand{\ee}{\end{itemize}}

\begin{document}
\title{Cooperation with Complement is Better}

\newcommand{\mypara}[1]{\vspace*{0.75ex}\noindent{\bf #1}~} 

\author{\.{I}lker Y{\i}ld{\i}r{\i}m  and Haluk Bingol\\
\institute{Department of Computer Engineering \\
 Bo\u{g}azi\c{c}i University\\
 Bebek, 34342, Istanbul, Turkey\\
 \tt{\{ilker.yildirim, bingol\}@boun.edu.tr}
}}
\date{}
\thispagestyle{plain}
\maketitle

\begin{abstract}
In a setting where heterogeneous agents interact to accomplish a given set of goals, cooperation is of utmost importance, especially when agents cannot achieve their individual goals by exclusive use of their own efforts~\cite{Castelfranchi1990}. Even when we consider friendly environments and benevolent agents, cooperation involves several issues: with whom to cooperate, reciprocation, how to address credit assignment and complex division of gains, etc.

We propose a model where heterogeneous agents cooperate by forming groups and  formation of larger groups is promoted. Benefit of agents is proportional to the performance and the size of the group. There is a time pressure to form a group. We investigate how preferring similar or complement agents in group formation affects an agent's success. Preferring complement in group formation is found to be better, yet there is no need to push the strategy to the extreme since the effect of complementing partners is saturated.

\end{abstract}
\section{Introduction}
\label{sec-intro}
In a setting where heterogeneous agents interact to accomplish a given set of goals, cooperation is of utmost importance, especially when agents cannot achieve their individual goals by exclusive use of their own efforts~\cite{Castelfranchi1990}. Even when we consider friendly environments and benevolent agents, cooperation involves several issues: with whom to cooperate, reciprocation, how to address credit assignment and complex division of gains, etc. Since we consider heterogeneous agents, one of the problems for an agent that considers to engage into cooperation is to decide a rule for choosing partners. Agents might prefer to cooperate with agents similar to themselves, or might prefer agents that are completely dissimilar from them, or they might even for groups composed of a mixture of several agent types.

Time is an important element of decision of who to cooperate. Given infinite time, one wants to get the perfect match. But if there is a time pressure, one makes the decision on what ever is available. Although one rejects an offer of a cooperation from a group at one time, the very same agent may accept the offer of the same group later in time.

One can also decide whom to interact and cooperate with by considering the trustworthiness of others. One of the most well known examples is Axelrod's iterated Prisoner's Dilemma~\cite{axelrod:1985}. Axelrod shows that the tit-for-tat strategy (cooperate in the first round, and do whatever the other agent does in the rest of the game) is the best within a controlled settings. But tit-for-tat is not the best strategy when one is playing with a selfish, hence a non-cooperative another one. In order to minimize losses, an adaptive trust strategy model is proposed by Wu and Sun~\cite{wusun2001}. 

Riolo, using the Iterated Prisoner's Dilemma studies  cooperation~\cite{riolo:1997} to show that tag mechanisms in the emergency of mutual cooperation is very useful. Edmonds also studies cooperation in symbiotic groups using agents with tags. In this model~\cite{Edmonds:2006}, agents who are able cooperate and form groups symbiotic groups can survive, because agents need all types of resources to survive but can only harvest one type of resource. But all groups are prone to invasion of selfish agents. Agents can cooperate only to others with close tag values, but the one among the close agents is selected randomly. In both models agents have no explicit preferences to select whom to cooperate. On the other hand, S. M. Deen and R. Jayousi assume knowledge of possible solutions and assume preferences among them~\cite{deenjayousi2006}. They setup a multiagent system in which each agent has its own preferences and those preferences may conflict. So they process preferences to match as much preferences as possible globally. 

In this study agents form groups to harvest fields. At the end of harvesting members of the groups get some payoff from the amount they collected and groups dissolve. This process repeats itself. The simulation is composed of terms. A {\it term} is composed of $T+4$ time units. What happens in one term briefly is:
\be
\item At $t=0$, among population $k$ of individuals are selected to form groups as {\it group initiators}. Each agent consumes a constant amount of food.
\item At $t=1$ to $t=T$, group initiators try to form their groups within $T$ time units. At each time unit each group initiator makes an offer to an agent that is not yet joined to a group. Note that if all the offers are granted, the maximum group size would be $T+1$.
\item At $t=T+1$, each group is assigned a field whose size is less then the total capacity of the group. While the group harvest the field, the reliability and efficiency of the agents take place.
\item At $t=T+2$, the collected food is shared. Agents in larger groups get better payoff. Therefore formation of larger groups is promoted. Friendship networks are updated according to the members of each group.
\item At $t=T+3$, all groups are dissolved. Group initiators are unassigned. 
\item After these time steps the next term starts.
\ee

In the rest of this paper, we explain parameters of an agent in Section~\ref{sec-model}. In Section~\ref{sec-oneturn} we describe one term in a simulation in detail and we explain how an offer is done and evaluated by agents. In Section~\ref{sec-simcomp} we describe different possible strategies an agent can have in choosing with whom to cooperate. The simulation setup is explained in  Section~\ref{sec-simsetup}. We give and discuss our results in Section~\ref{sec-conclusion}.

\section{Model Setup}
\label{sec-model}
The model is driven by this main assumption that there are many fields of different sizes such that individuals require to form groups in order to be able to harvest a field, even to harvest the field with the smallest size. In addition, it is assumed that groups do not need to compete for the same field, because there are many fields which are exactly the same. Thus individuals need to select partners to be in the same group among their friends (Each individual has friends which is a set of other individuals). Thus, the necessity to be in a group to be able to harvest food is the basic assumption. The environment is closed (there is no reproduction) on which we show that the preferences on selecting or forming a group, cooperation, affects the amount of food an individual can harvest. The simulation consists of time steps in which individuals go into groups, consume some food and may die if there is no food to consume.

Each agent has its 
\be
\item capacity, $c_i$, which is taken a constant, $c_0$ is this study;
\item reliability and efficiency vector $b_i$ = $[b_{i}^{r} b_{i}^{e}]^\tau$ 
where both $b_{i}^{r}$ and $b_{i}^{e}$ come from a random uniform distribution, $U(0,1)$ and $\tau$ is the transpose operation;
\item preference vector $p_i$ = $p_{i}^{r} p_{i}^{e}$ 
where  $p_{i}^{r}$ and $p_{i}^{e} \in [0,1]$ and $p_{i}^{r} + p_{i}^{e} = 1$;
\item savings of agent $a_i$ at time $t$, $s_{i}(t)$;
\item and a dynamic friend network of agent $a_i$ at time $t$, $F_{i}(t)$.
\ee

Population consists of $n$ agents, and no new agents are added to the population during the simulation. In each term $k$ of $n$ agents are randomly selected to form groups, where $k < n$. Fields are unlimited resources. Each group is assigned a field proportional to its size. We describe all of these parameters in more detail in section~\ref{sec-oneturn}. 

\section{Dynamics of one Term}
\label{sec-oneturn}
A term consists of $T+4$ time units. At time $t=0$, in each term $k$ of $n$ agents, the population, are randomly selected as the group initiators. The $k$ group initiators are responsible for forming groups in a limited time.

\subsection{Forming Groups}
\label{subsec-formgroups}
There is $T$ time units to form a group. The goal is to try to form a group with as large total capacity as possible. The group initiator has to have some tradeoff: It has to form the group within a given time and it wants to form that group based on its taste, that is vector $p_i$. 

Since each agent has a fixed and constant capacity of $c_0$ the largest group has also the largest capacity. At one unit time the initiator can make one offer to an agent. Since given $T$ time units, the maximum group size can be $T+1$. The agent $a_j$, who gets the offer may accept or reject the offer whose dynamics will be given later. The important point at this time is that similar to the group initiator, the agent who gets the offer has also $T$ time units to join a group. So it also has time pressure on it. 

\subsubsection{Making the Offer}
\label{subsec-makeoffer}
Each group initiator, agent $a_i$, evaluates all friends in its friend network according to its preference vector, $p_i$. If one of friends, agent $a_j$ has vector $b_j = [b_{j}^{r} b_{j}^{e}]^\tau$ and the group initiator has preference vector $p_{i} = [p_{i}^{r} p_{i}^{e}]$ then the metric for $a_j$ is $v_{ij}$ = 
$
\begin{bmatrix}
	p_{i}^{r} p_{i}^{e}
\end{bmatrix}
\begin{bmatrix}
	b_{j}^{r}\\
	b_{j}^{e}
\end{bmatrix}
$ = $p_{i}^{\tau} b_{j}$. Then for agent $a_i$ probability of selecting agent $a_j$ is given as

$$
\frac{v_{ij}^\gamma}{\sum_{k\in F_i}v_{ik}^\gamma} 
$$

\begin{figure}
\centering
\includegraphics[scale=0.50]{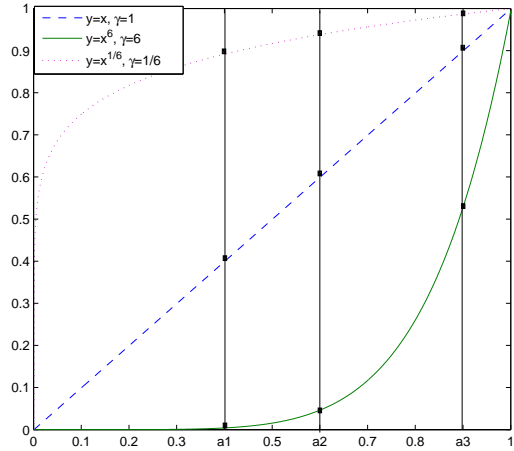}
\caption{Effect of $\gamma$}
\label{fig-offer}
\end{figure}

Note that $v_{ij}\in [0,1]$ and consider $\gamma$ power of $v_{ij}$, that is $e_{ij}^\gamma$. In Fig.~\ref{fig-offer}, there are agents $a_1$, $a_2$ and $a_3$. For $\gamma = 1$, $a_3$ will be selected instead of $a_2$. For $\gamma > 1$, the selection of $a_3$ becomes much cleaner, whereas for $\gamma < 1$ the difference of $a_2$ and $a_3$ decreases. In the extreme case $\gamma = 0$, the selection of $a_1$, $a_2$ and $a_3$ become the same. On the other hand, as $\gamma$ increases selection of $a_1$ and $a_2$ become close to each other and $a_3$ becomes more apart. 

We use the effect of $\gamma$ for {\it choosiness}. Although initially $a_3$ is very likely to get the offer, as time pressure increases, $a_2$ becomes considerable for the group. If time pressure is so high, even $a_1$ is also considered. 

Let 
$$
\gamma = \gamma_{0} \frac{T-t+|G(t)|}{T}
$$
where $t\in{1,...,T}$ is the time, and $T$ is the allowed time for group formation. $|G_{i}(t)|$ is the size of the group $i$ at time $t$. We take $\gamma_{0} = 3$ due to a similar model proposed by French and Kus in ~\cite{FrenchKus2008}. 

At $t=0$, $\gamma$ is high. As $t$ increases, the remaining time to form a group decreases. Then the pool of potential partners must be increased. This can be done by lowering $\gamma$. So that even $a_1$ in Figure~\ref{fig-offer} can get a chance. 

On the other hand, group size has the opposite effect. If group is large enough, it may not get a new agent that does not fit. But if the size is not large enough, obtaining a new agent should be promoted. Hence if the group size is low, $\gamma$ should be low, but if it is high, $\gamma$ need not be small.

\subsubsection{Evaluating the Offer}
\label{subsec-receiveoffer}
Agent evaluates the offer it gets based on all the offers it so far got. It  considers the time pressure. This time choosiness is defined as $ln(\gamma \frac{T-t}{T})$. 

An offer received from a group is evaluated based on the agents in the group that makes the offer. A vector which contains the mean of the reliability and that of efficiency is taken as the parameters of the group, 
$$
b_{G_{i}(t)} = 
\begin{bmatrix}
	b_{G_{i}(t)}^{r}\\
	b_{G_{i}(t)}^{e}
\end{bmatrix}
$$

Similar to the what group initiators do, 
$$
o_{ji}(t) = b_{G_{i}(t)}^{T} p_j
$$
 is calculated, which is the offer received by agent $a_j$ at time $t$ from group $i$ in a given term. The time pressure effect is included as 
$$
o_{ji}^{'}(t) = o_{ji}(t) + ln(\gamma_{0} \frac{T-t}{T})
$$
where $o_{ij}^{'}(t)$ is the adjusted value. The agent calculates the mean $\mu$ and standard deviation $\sigma$ of all the previous offers. If the adjusted value is larger than $\mu-1.5\sigma$ then the offer is accepted, otherwise it is rejected. Agent $a_j$ who received the offer, updates its $\mu$ and $\sigma$ for each offer it received.

Note that even if the group who made the offer is exactly the same, an agent may accept their second offer due to time pressure.

\subsection{Harvesting}
\label{subsec-harvesting}
After group formation, that $t=T+1$, a field is assigned to each group of size greater than $1$ which means in order to be assigned a field, the group should have at least two members. We assume that fields are not critical resources. Every group is assigned a field that is slightly less than the total capacity of the group. Since we assume that capacity of agents are the same, $c_0$, the size of the assigned field is as follows: 
$$
C_f = |G|c_{0} - |N(0, |G|)|
$$
where $N(0, |G|)$ is a normal distribution with mean $0$ and variance $|G|$, and $|G|$ is the size of the group. As the size of the group increases, fluctuation of the field size also increases. 

After group gets its field to harvest, every agent in the group harvest the field. An agent $a_j$ has capacity $c_0$, reliability $b_{j}^{r}$ and efficiency $b_{j}^{e}$. The amount it collects is given as:

\begin{equation*}
	A_{j} = 
	\begin{cases}
		c_{0} & \text{with probability $b_{j}^{r}$}\\
		c_{0} - |N(0,2)| & \text{with probability $1-b_{j}^{r}$}
	\end{cases}
\end{equation*}

Since agents are not efficient, the net amount is given as
$$
A_{j}^{n} = A_{j} b_{j}^{e}
$$

\subsection{Sharing The Collected and Friend Network Updates}
\label{subsec-sharing}
Since agent is in a group the amount it gets is proportional to its contribution to the group. In the case of single group, agent $a_j$ gets
$$
A_{j}^{n} \frac{A_{j}}{\sum_{k \in G_j}A_{k}}
$$
where $G_j$ is the group agent $a_j$ belongs to.
Since there are many groups and we want to favor larger group sizes, the actual award agent $a_j$ gets is:
\begin{equation*}
\Delta A_{j} = A_{j}^{n} \frac{A_{j}}{\sum_{k \in G_j}A_{k}} \left( 1 + \frac{|G_{j}|}{|G_{max}|}\right)
\end{equation*}
where $G_{j}$ is the group $a_j$ belongs to, $G_{max}$ is the largest possible group. The net gain $\Delta A_{j}$ is added to the savings $s_j$ of $a_j$.

If agents cannot gather all the field, because of they are not reliable enough, which means $\sum_{k \in G_{j}}A_{k} < C_f$, although they receive award for the amount of field they collected, they receive punishment for the remaining. The amount of punishment is evaluated using net amount collected by each agent. Each agent gets equal payoffs from the total punishment. The punishment agent $a_j$ receives is as follows:
$$
\frac{1}{|G_j|} \left(C_f - \sum_{k \in G_{j}}A_{k}^{n} \left(1 + \frac{|G_{j}|}{|G_{max}|}\right)\right) 
$$ 
where $G_{j}$ is the group $a_j$ belongs to, $G_{max}$ is the largest possible group. The savings $s_j$ of $a_j$ is decreased that much.

In each group, if there are pairs of agents which are not friends currently, these pairs may become friends. Such pairs occur when two agents are friends of the same group initiator, but they are not friends of each other. If there is such a pair in the group, each agent in such a pair firstly evaluates the value of the friend candidate using its preference vector, $p$. That agent also evaluates its current friends again by its preference vector, $p$. If the value of candidate agent is acceptable when compared to the current friends, the candidate is accepted, otherwise it is rejected. If both agents in the pair accept each other, then they become friends. Similar mechanism described in section~\ref{subsec-receiveoffer} is used to decide whether accept the candidate or reject it.

At the end of this term, at $t=T+3$, friendship network us updated as explained above. Cooperations (groups) resolve at the end of the term. A new term starts.

\section{Prefer the Similar or the Complement}
\label{sec-simcomp}
An agent, using its preference vector $p$, may prefer one individual to another or one group to another. Here we divide these preferences agents can have into two: 1) Similar Preferring, 2) Complement Preferring. Assume agent $a_j$ has reliability and efficiency vector $b_{j} = [b_{j}^{r} b_{j}^{e}]$ and preference vector $p_{j} = [p_{j}^{r} p_{j}^{e}]$. If $a_{j}$ is a {\it similar preferring} then it satisfies one of the following:
\begin{equation*}
p_{j}^{r} > p_{j}^{e} \text{ if } b_{j}^{r} > b_{j}^{e}
\end{equation*}
\begin{equation*}
p_{j}^{r} < p_{j}^{e} \text{ if } b_{j}^{r} < b_{j}^{e}
\end{equation*}

On the other hand, if $a_j$ is {\it complement preferring} then it satisfies one of the following:
\begin{equation*}
p_{j}^{r} > p_{j}^{e} \text{ if } b_{j}^{r} < b_{j}^{e} 
\end{equation*}
\begin{equation*}
p_{j}^{r} < p_{j}^{e} \text{ if } b_{j}^{r} > b_{j}^{e}
\end{equation*}

An agent may prefer some level of similarity or complementarity. This depends on its preference vector $p_j$. If $p_{j}^{r}$ and $p_{j}^{e}$ are close then it prefers agents whose reliability, $b_{j}^{r}$ and efficiency, $b_{j}^{e}$ are close. Similarly, if agent has $p_{j}^{r}$ and $p_{j}^{e}$ values far away, then it prefers the ones with reliability, $b_{j}^{r}$ and efficiency, $b_{j}^{e}$ apart. We use the parameter $\alpha$ to manage how apart $p_{j}^{r}$ and $p_{j}^{r}$ values of a preference vector $p_{j}$ will be. Take an agent $a_j$ who is complement preferring agent with $b_j = [b_{j}^{r} b_{j}^{e}]$ and $b_{j}^{r} > b_{j}^{r}$. Here is how we calculate preference vector, $p_j$:
$$
p_{j}^{r} = \frac{b_{j}^{e}}{b_{j}^{r}+b_{j}^{e}} - (1-b_{j}^{r})\alpha
$$
$$
p_{j}^{e} = \frac{b_{j}^{r}}{b_{j}^{r}+b_{j}^{e}} + (1+b_{j}^{r})\alpha
$$
where $\alpha$ to manage affect distance between preference on reliability, $p_{j}^{r}$ and preference on efficiency, $p_{j}^{r}$. Note that $p_{j}^{r} + p_{j}^{e} = 1$. The calculation is similar when $b_{j}^{r} < b_{j}^{e}$ or when the agent is a similar preferring agent.

The percentage of individuals that are complement preferring is again a parameter in the model, $\beta$. 

\begin{figure}
\centering
\includegraphics[scale=0.60]{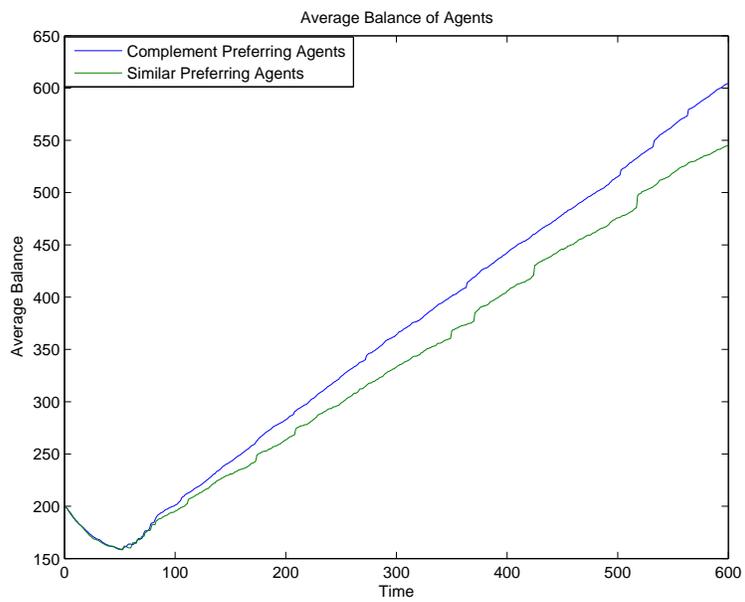}
\caption{Average food balance of individuals which prefer  complement and which prefer similar in $1$ run}
\label{fig-onerunaverage}
\end{figure}

\section{Simulation Setup}
\label{sec-simsetup}

The parameter setting of the simulations is as follows: The initial population size, {\it $n$} is $200$, the maximum number of terms in one simulation, {\it maxTerm} is $1000$, the number of group initiators in each term, {\it $k$} is $40$, the amount of initial food each agent has, {\it $s_{i}(0)$} is $200$, and the length of period of group formation, {\it $T$} is $8$. For simplicity, capacity of each agent is the same and set to $10$.

In Fig.~\ref{fig-onerunaverage}, the value of the parameter $\alpha$ is $1$, individuals prefer extremes. Percentage of complement preferring individuals, $\beta$ is $50\%$ in the population. The amount of food an agent consumes in one simulation step is $4$. The reason of decline in food balances in the beginning of the simulation is the sizes of friend networks of each individual is small in the beginning, hence they cannot form groups efficiently. But after the decline, we see that complement preferring agents achieve a higher food balance on the average when compared to the similar preferring agents. 

In Fig.~\ref{fig-simresults}, the amount of food an agent consumes in one simulation step, {\it ConsumptionAmount} is $2$. We run simulations for $45$ different configurations: $5$ different values of $\alpha$, $0.2$, $0.4$, $0.6$, $0.8$, $1.0$ and $9$ different values of percentage of complement agents, $\beta$, $10\%$, $20\%$, ..., $90\%$. For each configuration, we did $800$ replications over simulation runs of $1000$ terms. Hence the values on the $Y$ axis are the percentage of simulations in which complement preferring agents make more food balance than similar preferring agents. Since standard deviation is very small we ignore variance in our results.


\section{Conclusion}
\label{sec-conclusion}
\begin{figure}
\centering
\includegraphics[scale=0.60]{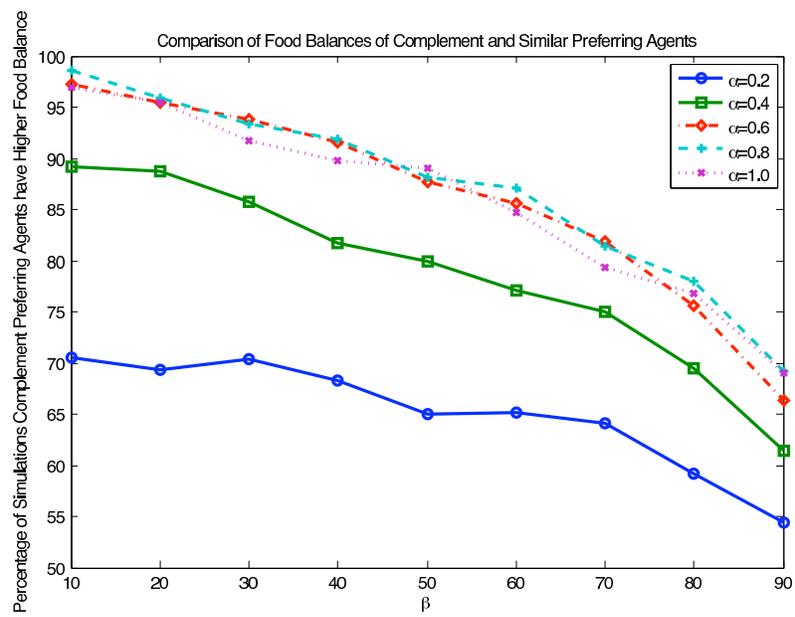}
\caption{Average number of times average food balances of complement preferring agents are higher than that of similar preferring agents over $800$ runs for each 45 different configurations}
\label{fig-simresults}
\end{figure}
We use the mate choice model introduced by French and Kus in~\cite{FrenchKus2008} with some changes. Group formation is actually a mate choice problem. In our model one agent is selecting many mates, partners, not one agent as in the mate choice problem. An agent who receives an offer should then consider all other partners in addition to the offer making agent which is the group initiator. In this study we make a description of that modified mate choice model, the group formation model in detail.

In Fig.~\ref{fig-simresults}, we see that for all values of $\alpha$ average number of times complement agents make more food balance decrease when percentage of complement preferring agents increase. Hence, if a small sub-population is complement then they achieve really better. This result is intuitive, since a small part of the population prefers the complementing partners, they are able to achieve cooperating with the complements. But when most of the agents want to cooperate with complements, complement for each agent may not be available.  

Secondly, for larger $\alpha$, we have better results, which implies that cooperating with more complement agents is better. But it does not change further after $\alpha$ is $0.6$. There is no need to push the strategy to the extreme since the effect of complementing partners is saturated after $\alpha$ exceeds $0.6$. 


Actually our findings are in compliance with that of~\cite{cowanjonard2004} in which Cowan and Jonard show that the diffusion of knowledge is best achieved when $10\%$ of the nodes in a group has weak links. In our study, we show that if $10\%$ of the population (set of all nodes in knowledge diffusion study) is complement preferring (has weak links) then they achieve the best depending on how complement agents (weak links) they are preferring.

\bibliographystyle{splncs} 
\bibliography{yildirim-CooperationWithComplementIsBetter} 
\end{document}